\documentclass[aps,prd,superscriptaddress,twocolumn,showpacs,nofootinbib]{revtex4-1}

\usepackage{amsmath}
\usepackage{amssymb}
\usepackage{graphicx}
\usepackage{mathrsfs}

\def\I{\mathrm{i}}
\def\vec#1{\mathbf{#1}}
\def\be{\begin{equation}}
\def\ee{\end{equation}}

\def\del#1#2{\frac{\partial #1}{\partial #2}} 
\def\D{\mathrm{d}}
\def\E{\mathrm{e}}
\def\I{\mathrm{i}}

\newcommand\hairspace{\kern .08333em }
\newcommand{\Real}{\Re\!\hairspace\mathfrak{e}\hairspace}
\newcommand{\Imag}{\Im\mathfrak{m}\hairspace}

\begin{document}

 \title{Quantum-gravitational effects on gauge-invariant scalar and tensor perturbations during inflation: The slow-roll approximation}
 
 \author{David Brizuela}
 \email{david.brizuela@ehu.eus}
 \affiliation{Fisika Teorikoa eta Zientziaren Historia Saila, UPV/EHU,
   644 P.K., 48080 Bilbao, Spain} 
 
 \author{Claus Kiefer}
 \email{kiefer@thp.uni-koeln.de}
 \affiliation{Institut f\"ur Theoretische Physik, Universit\"{a}t zu
   K\"{o}ln, Z\"{u}lpicher Stra\ss e 77, 50937 K\"{o}ln, Germany} 
 
 \author{Manuel Kr\"amer}
 \email{m.kraemer@wmf.univ.szczecin.pl}
 \affiliation{Instytut Fizyki, Uniwersytet Szczeci\'{n}ski,
   Wielkopolska 15, 70-451 Szczecin, Poland} 
 
 \date{\today}

\begin{abstract}
We continue our study on corrections from canonical
quantum gravity to the power spectra of gauge-invariant inflationary
scalar and tensor perturbations. A direct canonical quantization of
a perturbed inflationary universe model is implemented,
which leads to a Wheeler-DeWitt equation. For this equation, a semiclassical
approximation is applied in order to obtain a Schr\"odinger equation with
quantum-gravitational correction terms, from which we calculate the
corrections to the power spectra. We go beyond the de
Sitter case discussed earlier and analyze our model in the first
slow-roll approximation, considering terms linear in the slow-roll
parameters. We find that the dominant correction term from the de
Sitter case, which leads to an enhancement of power on the largest
scales, gets modified by terms proportional to the slow-roll
parameters. A correction to the tensor-to-scalar
ratio is also found at second order in the slow-roll parameters.
Making use of the available experimental data, the magnitude
of these quantum-gravitational corrections is estimated. Finally,
the effects for the temperature anisotropies in the cosmic microwave
background are qualitatively obtained.
\end{abstract}

 \maketitle

\section{Introduction}

The search for a theory of quantum gravity will only be successful if
one eventually finds a way to test the candidate theories by
experiment or observation \cite{oup,KK12}. Because of the extremely
high energies at which quantum-gravitational effects are expected to 
be strong, many researchers have looked for features of
quantum-gravitational origin in the anisotropies of the cosmic
microwave background (CMB); see, for example,
\cite{KK-PRL, BEKKP,Gianluca,BE14,ACT06,Sasha1,Sasha2,Sasha3,KTV16,AAN,MO16,BGS15,SBB16,AG16}.
These anisotropies are thought to originate during the 
inflationary phase of the primordial universe from quantum
fluctuations of the metric and the scalar inflaton
field. Hence, they are in a sense already a consequence of the quantum
nature of space-time and thus an effect of quantum gravity. What we
are going to study in the following 
is to derive the corrections that arise from quantizing the
Universe as a whole in the canonical formalism that leads to the
Wheeler-DeWitt equation. 

One way to obtain corrections to the power spectra of the inflationary
scalar and tensor perturbations, which lead to the CMB anisotropies, is
to perform a semiclassical approximation of the Wheeler-DeWitt
equation \cite{singh}. This procedure leads to a Schr\"odinger
equation with quantum-gravitational correction terms, which can be
used to calculate the corrected power spectra. This was analyzed for a
simple model containing only non-gauge-invariant perturbations of a
scalar field in \cite{KK-PRL,BEKKP,Gianluca}. An alternative
semiclassical approximation was presented in \cite{bertoni} and was
used to calculate CMB corrections in
\cite{BE14,ACT06,Sasha1,Sasha2,Sasha3,KTV16}. The results of both approaches
were essentially the same, with differences only in the numerical
factors. 

 We have previously studied the corrections to the power spectra of
 gauge-invariant scalar and tensor perturbations and have made
 explicit calculations for the de~Sitter case \cite{BKK}. Here, we
 shall extend our analysis and use a {\em generic slow-roll approximation}
 that is compatible with all observational data obtained so far and
 which encompasses a wide range of inflaton potentials. 

Our paper is organized as follows. In Sec.~II, we summarize
the quantization procedure and the semiclassical
approximation. This leads to the equations we use in Sec.~III to
obtain general 
expressions for the power spectra without and with quantum gravity
corrections. In Sec.~IV, we introduce the slow-roll
approximation, and in Sec.~V we calculate the corresponding
uncorrected power spectra. Sec.~VI represents the main
part of the paper; here, we calculate the slow-roll power spectra with
the quantum-gravitational corrections. In Sec.~VII, we derive 
from this the corrections to
the spectral index, its running, and to the tensor-to-scalar ratio.  
We also comment on the observability of the calculated effects
and compute the correction for the CMB temperature anisotropies.
Finally, in Sec.~VIII, we summarize and perform an outlook. We also
compare our results with the results obtained in \cite{KTV16}.

\section{Quantization and semiclassical approximation 
  of a perturbed inflationary universe} 

Here and in the next section, we present a brief self-contained
summary of our earlier results from \cite{BKK}. 
We model an inflationary universe as a flat
Friedmann-Lema\^itre-Robertson-Walker (FLRW) space-time
plus fluctuations. The matter content is supposed to be a massive scalar inflaton field
$\phi$ with potential $\mathcal{V}(\phi)$, and a vanishing cosmological
constant is assumed. For convenience, we will use conformal time 
$\eta$, which is defined in terms of the cosmic time $t$ and the scale
factor $a(t)$ by ${\D\eta}/{\D t} = a^{-1}$. Furthermore,
introducing four space-time functions $A$, $B$, $\psi$, $E$ to describe
the scalar perturbations, as well as the symmetric spatial tensor $h_{ij}$ to
encode the tensorial degrees of freedom, the metric of the FLRW space-time 
plus fluctuations reads (see, for example, \cite{Mukhanov:1992,PU09})
\begin{align}
\label{metric}
\D s^2=a^2(\eta)\Bigl\lbrace
&-\left(1-2A\right)\D\eta^2
+2\left(\partial_iB\right)\D x^i\D \eta \\
&+\left[\left(1-2\psi\right)\delta_{ij}
+2\partial_i\partial_jE + h_{ij}
\right]\D x^i\D x^j\Bigr\rbrace\,. \nonumber
\end{align}
The additional scalar perturbations $\varphi(\eta,\vec{x})$ of the
field $\phi$ can be combined with the scalar metric perturbations to construct
a master gauge-invariant quantity called the \emph{Mukhanov-Sasaki variable}
\cite{Mukhanov:1992,PU09}:
\be \label{Mukhdef} 
v(\eta,\vec{x}):=a
\left[\varphi+\frac{\phi^\prime}{\mathscr{H}}\left(A+2\,\mathscr{H}\left(B-E^\prime\right)
    + \left[B-E^\prime\right]^\prime\right)\right], 
\ee
where we have indicated the derivative with respect to $\eta$ with a
prime and used the definition $\mathscr{H} := a'/a$.

We denote the Fourier transform of \eqref{Mukhdef} as $v_{\vec{k}}$
and also introduce the Fourier-transformed perturbation variable of
the gauge-invariant tensor perturbations $h_{ij}$ with polarization
$\lambda \in \{+,\times\}$ as 
\be 
v_{\vec{k}}^{(\lambda)} := \frac{a\,h^{(\lambda)}_{\vec{k}}}{\sqrt{16\pi G}}.
\ee 
We note that an important feature in the definition of these variables
is the rescaling with respect to $a$.

The action $S$ for our perturbed inflationary universe then takes the form 
\begin{align} \label{genaction}
S = \frac{1}{2}\int
\D\eta\,\Biggl\lbrace\,\mathfrak{L}^3&\left[-\,\frac{3}{4\pi G}\,(a')^2 +
  a^2\,(\phi')^2-2\,a^4\,\mathcal{V}(\phi)\right] \nonumber \\
  +\,\frac{1}{\mathfrak{L}^3}\sum_{\vec{k}}
&\,\Bigl[ v_{\vec{k}}^\prime{v_{\vec{k}}^*}^\prime
+{}^\text{S}\omega^2_{\vec{k}}\,v_{\vec{k}}v_{\vec{k}}^*\Bigr] \Biggr. \nonumber\\
+\,\frac{1}{\mathfrak{L}^3}\sum_{\lambda=+,\times}\sum_{\vec{k}}
&\left[ v_{\vec{k}}^{(\lambda)\prime}v_{\vec{k}}^{(\lambda)*\prime}
+{}^\text{T}\omega^2_{\vec{k}}\,v_{\vec{k}}^{(\lambda)}v_{\vec{k}}^{(\lambda)*}\right]\Biggr\rbrace\,.
\end{align}
Here, we have introduced the ``frequencies'' ${}^\text{S}\omega$ and 
${}^\text{T}\omega$ by
\be
\label{defomega}
{}^\text{S}\omega^2_{\vec{k}}(\eta):=k^2-\frac{z^{\prime\prime}}{z}\,,\quad
{}^\text{T}\omega^2_{\vec{k}}(\eta):=k^2-\frac{a^{\prime\prime}}{a}\,,
\ee
where $z$ is defined as $z:=a\,\phi^\prime/\mathscr{H}$.

In order to avoid the infinity arising from the volume integral of the action,
it is necessary to introduce a maximum length scale $\mathfrak{L}$,
which can be thought of as the maximum size of the universe under consideration
(see e.g. the Appendix of \cite{Parker:2009}). 
We will later have to specify a value for
$\mathfrak{L}$ when comparing our results to observations; however, up
to that point it is possible to remove $\mathfrak{L}$ from the notation by
applying the following redefinitions \cite{Sasha2,BKK}: 
\be \label{repl1}
a_\text{new} = a_\text{old}\,\mathfrak{L}\,, \; \eta_\text{new} =
\frac{\eta_\text{old}}{\mathfrak{L}}\,,
\; k_\text{new} = k_\text{old}\,\mathfrak{L}  \,,\; v_\text{new} = \frac{v_\text{old}}{\mathfrak{L}^2}\,.
\ee
Note that, after these rescalings, $a$ obtains the dimension of a
length, whereas $\eta$, $k$, and $v_{\vec{k}}$ are dimensionless. 

In order to perform an entirely consistent quantization, one should use a real set of variables instead of the 
complex $v_{\vec k}$ variables. Such real variables can be constructed from a double copy of the complex variables and the wave function,
see, for instance, \cite{Martin:2012}.
Nonetheless, since such a redefinition will not influence our calculations later on, we will not introduce these new variables and thus treat $v_{\vec k}$ as if it were real in order to keep our presentation brief and concise.

After a canonical quantization \cite{oup} of \eqref{genaction},
and choosing a product ansatz for the full wave function,
we end up with a Wheeler-DeWitt equation of the following form,
for \emph{each} mode $\vec{k}$ and for \emph{both} the scalar
and tensor perturbations,
\begin{align}
\label{WdWmaster}
\frac{1}{2}\Biggl\{\E^{-2\alpha}\biggl[&\frac{1}{m_\text{P}^2}\,\del{^2}{\alpha^2}-\del{^2}{\phi^2}+2\,\E^{6\alpha}\,\mathcal{V}(\phi)\biggr] \nonumber \\
&-\,\frac{\partial^2}{\partial v_{\vec k}^2}+\omega^2_{\vec
  k}(\eta)\,v_{\vec{k}}^2\Biggr\}\Psi_{\vec{k}}(\alpha,\phi,v_{\vec{k}})
= 0\,. 
\end{align}
We have combined the notation for the scalar and tensor modes and thus
removed the superscripts S, T and $(\lambda)$. We also have set 
(besides $c=1$) $\hbar=1$ and introduced the dimensionless quantity $\alpha :=
\ln\!\left(a/a_0\right)$, where $a_0$ is a reference scale factor.
For simplicity, we will not write out this reference scale in the
following (or, equivalently, $a_0=1$ will be imposed); but it is
implicitly understood that a factor $a_0$ is associated with every
factor of $\E^{\alpha}$. We have also defined a rescaled Planck
mass $m_\text{P}$ in order to incorporate several numerical prefactors, 
\be \label{mpdef}
m_\text{P}^2 := \frac{3}{4\pi G}\,.
\ee
We shall use $m_\text{P}^2$ as the parameter with respect to which 
the semiclassical approximation of \eqref{WdWmaster} is carried out
\cite{singh,KK-PRL}. For this purpose, $\phi$ has to be rescaled to the
  dimensionless variable
  \be
  \tilde{\phi} := m_\mathrm{P}^{-1} \phi.
  \ee
More precisely, the reason behind this rescaling is that
we will subsequently perform an expansion in inverse powers of $m_\mathrm{P}$
and, in order to obtain the correct classical background equations at
first order in that expansion, it is necessary that the first
two terms in square brackets of \eqref{WdWmaster} have the same power
of $m_\mathrm{P}$. In any case, once the expansion is made,
we will revert and provide all expressions in terms of $\phi$.
  
The approximation is then performed by expanding the wave function
$\Psi_{\vec{k}}$ using the functions
$S_i(\alpha,\tilde\phi,v_{\vec{k}})$, $i = 0,1,2,\dots$, as follows: 
\be \label{WKBansatz}
\Psi_{\vec{k}} = \exp\Bigl[\I\left(m_{\rm P}^2\,S_0 + m_{\rm P}^0\,S_1 + m_{\rm
  P}^{-2}\,S_2 + \ldots\right)\Bigr].
\ee
After inserting this WKB-type ansatz into \eqref{WdWmaster}, the terms
containing a certain power of $m_{\rm P}$ are collected and their sum
is set equal to zero.  

At order $m_\text{P}^2$, we obtain the Hamilton-Jacobi equation of the
minisuperspace background,
\be \label{HJeq}
-\left(\del{S_0}{\alpha}\right)^2 + m_{\rm
  P}^2\left(\del{S_0}{\phi}\right)^2+\frac{2\,\E^{6\alpha}}{m_{\rm
    P}^2}\,\mathcal{V}(\phi)=0\,. 
\ee
The Planck mass occurs explicitly here because the expansion in
\eqref{WdWmaster} is performed in the form of the equation using
$\tilde\phi$ instead of $\phi$. 
 
At the next order, $m_\mathrm{P}^0$, it is possible to write the corresponding
equation for $S_1$ as a Schr\"odinger equation for a related wave function
$\psi^{(0)}_{\vec{k}}$ in the following way,
\be \label{Schreq}
\mathcal{H}_{\vec{k}}
\psi^{(0)}_{\vec{k}}=\I\,\del{}{\eta}\,\psi^{(0)}_{\vec{k}},
\ee
where the conformal time is defined in terms of the minisuperspace
variables by 
\be \label{ctimedef}
\del{}{\eta} :=
\E^{-2\alpha}\left[-\,\del{S_0}{\alpha}\del{}{\alpha}+m_{\rm
    P}^2\,\del{S_0}{\phi}\del{}{\phi}\right], 
\ee
and the perturbative Hamiltonian operator is given as follows,
\begin{align}
\mathcal{H}_{\vec{k}} :=-\,\frac{1}{2}\,\frac{\partial^2}{\partial
  v_{\vec k}^2}+\frac{1}{2}\,\omega^2_{\vec
  k}(\eta)\,v_{\vec{k}}^2. \label{SEqoneside} 
\end{align}
The information at the next order, $m_\mathrm{P}^{-2}$,
can be encoded in a wave function $\psi^{(1)}_{\vec{k}}$,
related to the function $S_2$, which obeys the following corrected
Schr\"odinger equation:
\begin{align} \label{corrSchreq} 
\I\,\frac{\partial}{\partial \eta}\,\psi^{(1)}_{\vec{k}} =
\mathcal{H}_{\vec{k}}\psi^{(1)}_{\vec{k}}
-\frac{\psi^{(1)}_{\vec{\vec{k}}}}{2\,m_{\rm P}^2
  \,\psi^{(0)}_{\vec{k}}}&\Biggl[\frac{\bigl( 
\mathcal{H}_{\vec{k}}\bigr)^2}{V}\,\psi^{(0)}_{\vec{k}} \\
&+ \I\,\frac{\partial}{\partial \eta}\!
\left(\frac{\mathcal{H}_{\vec{\vec{k}}}}{V}\right)
\psi^{(0)}_{\vec{k}}\Biggr]\,. \nonumber
\end{align}
Here, we have defined an auxiliary potential as
\be \label{potential}
V(a,\phi) := \frac{2\,a^4}{m_{\rm P}^2}\,\mathcal{V}(\phi),
\ee
which has the dimension of a length squared.

In summary, the most important relations of this section are the
two wave equations \eqref{Schreq} and \eqref{corrSchreq}.
The former one describes quantum fluctuations evolving on
a classical (FLRW) background space-time, while the latter
one encodes also corrections arising from the quantum
behavior of that background. The aim of this paper is thus
to solve these two equations with relevant initial conditions in order
to obtain the power spectra with quantum-gravitational corrections
for inflationary perturbations. In the next section, the
ansatz to be used for such a purpose is described, as well
as the explicit form of the power spectra.

\section{Gaussian ansatz and expressions for the power spectra}
\label{sectPS}

We use a Gaussian ansatz for both the uncorrected Schr\"odinger
equation \eqref{Schreq} and the corrected one
\eqref{corrSchreq}, with the normalization factor
$N_{\vec{k}}^{(0,1)}(\eta)$ and the inverse Gaussian widths 
$\Omega_{\vec{k}}^{(0,1)}(\eta)$ \cite{BKK},
\be
\label{Gaussianansatz}
\psi^{(0,1)}_{{\vec{k}}}(\eta,v_{\vec{k}}) =
N_{\vec{k}}^{(0,1)}(\eta)\,\text{e}^{-\frac{1}{2}\,\Omega_{\vec{k}}^{(0,1)}(\eta)\,v_{\vec{k}}^2}. 
\ee
Here, and in the following, the superscript $(0)$ stands for the uncorrected and $(1)$
for the corrected case. As will be shown below,
in order to obtain the power spectra for the scalar and tensor
perturbations, we have to find the solutions for
$\Omega_{\vec{k}}^{(0,1)}$. For the uncorrected Schr\"odinger
equation, we have to solve 
\be
\I\,\Omega_{\vec{k}}^{(0)\prime}(\eta) =
\bigl(\Omega_{\vec{k}}^{(0)}(\eta)\bigr)^2-\omega^2_{\vec
  k}(\eta)\,, \label{eqomega0} 
\ee
while for the corrected Schr\"odinger equation, we have to find a solution to
\be
\I\,\Omega_{\vec{k}}^{(1)\prime}(\eta) =
\bigl(\Omega_{\vec{k}}^{(1)}(\eta)\bigr)^2-\widetilde\omega^2_{\vec
  k}(\eta). \label{Omegaeq} 
\ee
In this last expression the corrected frequencies $\widetilde\omega_{\vec{k}}$
have been defined as follows: 
\begin{eqnarray}\label{defcorrectedomega}
\widetilde\omega_{\vec{k}}^2:=\omega_{\vec{k}}^2-
\frac{1}{2 m_\mathrm{P}^2 V}\,
\Real\biggl[&\left(3\Omega^{(0)}_{\vec{k}}-{\rm i}\,(\ln V)'\right) \left(\omega^2_{\vec{k}} - (\Omega^{(0)}_{\vec{k}})^2\right) \nonumber \\
&+2\, \I \,\omega_{\vec{k}} \, \omega'_{\vec{k}}\biggr].
\end{eqnarray}
As is explicitly written in this definition, in this paper we will only consider
the real part of the correction terms for the frequencies, since the imaginary part
tends to exhibit unphysical behavior, in particular unitarity violation.
The presence of this term can be traced back to the fact that the starting point is the Wheeler-DeWitt equation, not the Schr\"odinger equation. Its treatment is subtle and not fully understood. 
Such a term can be relevant for calculating the probability of cosmological instabilities \cite{deSitter}, but we do not expect it to play a role for the corrections of the power spectrum. For a more
detailed discussion about this point we refer the reader to \cite{BKK}. 

Since the equation for $\Omega^{(1)}_{\vec k}$ is nonlinear, it is reasonable
to linearize it around the background solution $\Omega^{(0)}_{\vec k}$ 
by defining $\widetilde\Omega^{(1)}_{\vec k}:=\Omega^{(1)}_{\vec
  k}-\Omega^{(0)}_{\vec k}$. After neglecting 
the quadratic term, one obtains the following linear equation:
\begin{eqnarray}
\label{Omega1tildeq}
{\rm i}\, \widetilde\Omega^{(1)\prime}_{\vec k}&=&2\,
\Omega^{(0)}_{\vec k} \widetilde\Omega^{(1)}_{\vec
  k}-\left(\widetilde\omega^2_{\vec k}-\omega^2_{\vec k}\right). 
\end{eqnarray}
Since it is usually possible to find an analytical solution for $\Omega^{(0)}_{\vec k}$,
one is then left with only a linear equation for $\widetilde\Omega^{(1)}_{\vec k}$,
which is easier to solve than the full non-linear equation \eqref{Omegaeq}.

Apart from the equations themselves, another important key aspect of the analysis
is to choose appropriate and meaningful initial conditions. For the uncorrected case,
as it is usually done in quantum field theory, the Bunch-Davies vacuum will be chosen,
which in this setting means $\Omega^{(1)}_{{\vec k}}=k$.
The case of an initial non-Bunch-Davies state was recently discussed
in \cite{Broy16}. Moreover, as we have discussed in \cite{BKK}, the most
natural initial conditions for the corrected
case should be those that best resemble the properties of the Bunch-Davies vacuum.
Namely, when the mode is well inside its Hubble radius, it should be oscillating
with a fixed constant frequency and amplitude. This is achieved by
choosing
\begin{eqnarray}
\Real(\Omega^{(1)}_{{\vec k}})^2&=&\Real(\widetilde\omega^2_{{\vec k}}),\nonumber\\
\Imag (\Omega^{(1)}_{{\vec k}})&=& 0. \label{initialdata}
\end{eqnarray}
Since the imaginary part of the corrected frequencies has been neglected, these two relations
can simply be written as $\Omega^{(1)}_{{\vec k}}=\widetilde\omega_{{\vec k}}$.
For the linearized function the above conditions are rewritten as $\widetilde\Omega^{(1)}_{{\vec k}}=k-\widetilde\omega_{{\vec k}}$.
Therefore, the latter is the condition that will be used as initial data at early times ($\eta \rightarrow -\infty$).

The power spectrum for the scalar perturbations including the quantum-gravitational corrections is given by
\be \label{PS0omega}
{\cal P}_{\text{S}}^{(1)}(k) =
 \frac{4\pi G}{a^2\,\epsilon}\,\frac{k^3}{2\pi^2}\,\frac{1}{2\,\Real {}^{^\text{S}\!}\Omega_{\vec k}^{(1)}}\approx
 {\cal P}_{\text{S}}^{(0)}(k)\Big\{1+\Delta_{\text{S}}\Big\}\,, 
\ee
where $\epsilon$ is the slow-roll parameter defined below in 
\eqref{defepsilon}, and ${\cal P}_{\text{S}}^{(0)}(k)$ is the usual scalar power
spectrum defined by inserting $\Omega^{(0)}_{\vec k}$ instead of
$\Omega^{(1)}_{\vec k}$ into this expression. The quantum-gravitational effects are thus encoded
in the term 
\be\label{defdeltas}
\Delta_{\text{S}}:=-\,\frac{\Real{}^{^\text{S}\!}\widetilde\Omega^{(1)}_{\vec
    k}}{\Real{}^{^\text{S}\!}\Omega^{(0)}_{\vec k}}. 
\ee
This ratio has to be computed in the limit of super-Hubble scales
(or late times), given by $k\eta \rightarrow 0^-$,
when the perturbations get ``frozen''.

The power spectrum for the tensor perturbations is, consequently, obtained by
\be
\label{PT0omega}
{\cal P}_{\text{T}}^{(1)}(k) = \,\frac{64\pi G}{a^2}\,\frac{k^3}{2\pi^2}\,\frac{1}{2\,\Real {}^{^\text{T}\!}\Omega^{(1)}_{\vec k}}
\approx {\cal P}_{\text{T}}^{(0)}(k)\Big\{1+\Delta_{\text{T}}\Big\}\,.
\ee
Here, the same comment as in the scalar case about taking the super-Hubble scale limit for $\Real \Omega_{\vec k}$ applies, and the corrections are given by the term
\be
\Delta_\text{T}:=-\frac{\Real {}^{^\text{T}\!}\widetilde\Omega^{(1)}_{\vec k}}{\Real{}^{^\text{T}\!}\Omega^{(0)}_{\vec k}}.
\ee
Finally, the corrected tensor-to-scalar ratio $r^{(1)}$ is defined as
\be \label{tts0}
r^{(1)} := \frac{{\cal P}_{\text{T}}^{(1)}(k)}{{\cal P}_{\text{S}}^{(1)}(k)}\approx r^{(0)} \left(1+\Delta_\text{T}-\Delta_\text{S}\right),
\ee
where $r^{(0)}$ is the ratio found at the order of approximation that corresponds to quantum field theory on a fixed, curved background.
In the de~Sitter case, since ${}^\text{S}\widetilde\omega_{\vec k}={}^\text{T}\widetilde\omega_{\vec k}$, both corrections are the same,
$\Delta_\text{T}=\Delta_\text{S}$, and thus in our previous work \cite{BKK} 
there appeared no correction to this quantity at the
considered order of approximation. This will be different here.

To recap, note that the task of obtaining quantum-gravitational corrections to the scalar and
tensor power spectra reduces to solving equation \eqref{Omega1tildeq} with initial data
\eqref{initialdata}. Once this is done, one only needs to compute the corresponding
power spectra using the relations \eqref{PS0omega} and \eqref{PT0omega}.

\section{The slow-roll approximation}

In this section the well-known slow-roll approximation is described in
order to clarify and set up the notation. In addition, several
physical quantities, which appear in the key equations of motion \eqref{eqomega0}
and \eqref{Omega1tildeq}, will be explicitly given up to linear order in the slow-roll parameters, which we will define below.

Using the Hubble parameter $H = \dot{a}/{a}$, we can define the
\emph{first slow-roll parameters} $\epsilon$ and $\delta$ (see
e.g. Eqs (8.37) and (8.38) in \cite{PU09}),
\begin{align} \label{defepsilon}
\epsilon &:= -\,\frac{\dot{H}}{H^2} = 1- \frac{\mathscr{H}'}{\mathscr{H}^2}\,,\\
\delta &:= \epsilon - \frac{\dot{\epsilon}}{2H\epsilon} = 
-\,\frac{\ddot{\phi}}{H\dot{\phi}}\,. 
\end{align}
In terms of the slow-roll parameter $\epsilon$, the
equation $\dot{a}=Ha$ and the Hamiltonian constraint equation can be
written as follows: 
\begin{align} 
\frac{\ddot{a}}{a}&=(1-\epsilon) \,H^2\,,\\\label{hamiltonianeq}
H^2\left(1-\frac{\epsilon}{3}\right)&=\frac{2}{m_\text{P}^2}\,{\cal V}(\phi),
\end{align}
or, using the conformal time $\eta$,
\begin{align}
\frac{a''}{a}&=(2-\epsilon) \,{\mathscr H}^2\,,\\
{\mathscr
  H}^2\left(1-\frac{\epsilon}{3}\right)&=\frac{2a^2}{m_\text{P}^2}\,{\cal
  V}(\phi). 
\end{align}
No approximation has been assumed yet, so these relations are exact.

The first order of our approximation scheme employs the function $S_0$, which satisfies the Hamilton-Jacobi equation \eqref{HJeq} of the background.
One cannot write down, of course, a general solution of \eqref{HJeq} for general $\epsilon$ and thus for general potential ${\mathcal V}(\phi)$. 
But it is possible to recover \eqref{hamiltonianeq} from the Hamilton-Jacobi equation \eqref{HJeq} in the slow-roll approximation, in which terms
quadratic and higher of $\epsilon$ and $\delta$ are neglected. At this level of approximation, we find
\be
S_0 =
-\,\frac{a^3}{m_\text{P}}\,\sqrt{\frac{2}{3}\,
\frac{\mathcal{V}(\phi)}{3-\epsilon(\phi)}},  
\ee
where
\be
\epsilon(\phi)=\frac{m_{\rm P}^2}{12{\mathcal V}^2}\left(\frac{\partial{\mathcal V}}{\partial\phi}\right)^2.
\ee
This is a well-known expression for $\epsilon$; see, for example, Eq.~(8.49) in \cite{PU09}.\footnote{The different prefactor in this expression originates from our definition \eqref{mpdef} of the rescaled Planck mass $m_\text{P}$.}
Note that the sign of $S_0$ is not fixed by \eqref{HJeq}, and it is chosen such that the flux of time \eqref{ctimedef}
points in the direction of the expansion of the universe.

Let us now introduce in more detail the slow-roll approximation. It essentially consists in assuming 
\be
\dot{\phi}^2 \ll {\mathcal V}\,,\quad \ddot{\phi} \ll 3H\dot{\phi}\,.
\ee
It can be shown that this is equivalent to assuming that the slow-roll parameters $\epsilon$ and $\delta$
are small: $\epsilon\ll 1$ and $|\delta| \ll 1$. If one assumes those to
be vanishing, one would recover the de Sitter case analyzed in our
previous paper \cite{BKK}. Since the pure de Sitter case is often too restrictive to make
contact with observations, we shall now consider the first-order
slow-roll approximation, which assumes $\epsilon$ and $\delta$ to be small
but nonvanishing, and drop quadratic terms in those. 

In order to solve the equations presented in the previous section in
this approximation, we need to express the frequencies \eqref{defomega}
and \eqref{defcorrectedomega}, as well as the potential \eqref{potential} in terms of the
conformal time $\eta$ and the slow-roll parameters $\epsilon$ and $\delta$.

It is well known (see e.g.\,\cite{PU09}) that, in this approximation,\footnote{Note in this context that the quantity $z$ used in \eqref{defomega} can also be written as $z = a\sqrt{\epsilon}$.} the frequencies of the modes can be written as
\begin{eqnarray}\label{omegasgamma}
{}^\text{S}\omega^2_{\vec{k}}(\eta)&=& k^2 - \frac{2 +
  3\gamma}{\eta^2} + {\cal O}(2),\\\label{omegatepsilon} 
{}^\text{T}\omega^2_{\vec{k}}(\eta)&=& k^2 - \frac{2 +
  3\epsilon}{\eta^2}+ {\cal O}(2), 
\end{eqnarray}
where ${\cal O}(2)$ stands for terms quadratic in $\epsilon$ and
$\delta$ and where we have defined, for later convenience, 
the combination 
\be \label{gammadef}
\gamma:=2\epsilon-\delta.
\ee
Note that setting $\delta = \epsilon$, or equivalently
$\gamma=\epsilon$, converts the equation for scalar perturbations into
the one for tensor perturbations. Therefore we will perform
all calculations only for the scalar perturbations and obtain
the results for the tensor perturbations using this relation at the end.   

Making use of the Hamiltonian constraint as written in \eqref{hamiltonianeq}, it is possible to write the rescaled potential
\eqref{potential} as 
\be
V=a^4 H^2\left(1-\frac{\epsilon}{3}\right).
\ee
In order to write this expression explicitly in terms of the conformal time and the slow-roll parameters,
we use the definition of the Hubble factor in terms of the conformal time to write
\be
\eta=\int\frac{\D a}{a^2 H}.
\ee
Integrating this relation by parts twice, we get
\be\label{timeah}
\eta=-\frac{(1+\epsilon)}{a H}+{\cal O}(2).
\ee
Now we can express the Hubble parameter as $H=a'/a^2$ and integrate
this equation, which leads to 
\be
a=\frac{C}{(-\eta)^{1+\epsilon}}+{\cal O}(2),
\ee
with an integration constant $C$. The Hubble parameter then takes the form
\be\label{Hubbleeta}
H=H_{0} \left(\frac{\eta}{\eta_{0}}\right)^\epsilon+{\cal
  O}(2)=H_{0}\left[1+\epsilon\ln\!\left(\frac{\eta}{\eta_{0}}\right)\right]+{\cal
  O}(2), 
\ee
in which the constant $C$ has been replaced by
\be
C = (1+\epsilon)\,\frac{(-\eta_{0})^\epsilon}{H_{0}};
\ee
$H_{0}$ being the value of the Hubble parameter at
$\eta=\eta_{0}$. Physically, $H_{0}$ defines 
the de~Sitter space-time we are considering as our reference.
Below we will have to choose a value for $H_{0}$, since
the results will depend on it.
The only physically meaningful point that exists in the evolution
of different modes, is when they cross the Hubble horizon.
Therefore, we will choose $H_{0}$ as the value of the Hubble
factor at the Hubble-scale exit, that is, $H_{0}=H_k$, with
$H_k=k/a$. Following relation
\eqref{timeah}, this implies choosing $\eta_{0}=-1/k$ on the
right-hand side of \eqref{Hubbleeta}. 
(Note that in this equation, at the level of approximation chosen,
$\eta_{0}$ is only 
meaningful at order $\epsilon^0$.)  This choice makes
both $H_{0}$ and $\eta_0$ become $k$-dependent but, since the
evolution of every $k$-mode is independent, there is no problem
in assuming that they experience a different reference de~Sitter space-time.

As a side remark, note that the present
approximation is valid during times $\eta$ which obey
\be
|\eta_{0}|\E^{-1/\epsilon}\ll |\eta| \ll |\eta_{0}|\E^{1/\epsilon}.
\ee
Due to the smallness of $\epsilon$, this time range is very large. Nonetheless, when
considering the asymptotic value for different physical objects as $\eta\rightarrow-\infty$,
one should take into account that in this limit the approximation breaks down.

Finally, making use of all above results, the potential $V$ can be written
in the following way:
\begin{eqnarray} \label{Veps}
V&=&\frac{(a
  H)^4}{H^2}\left(1-\frac{\epsilon}{3}\right)\\
  &=&\frac{1}{H_k^2\eta^{4} (k\eta)^{2\epsilon}}\left(1+\frac{11\epsilon}{3}\right)+{\cal
  O}(2), \nonumber
\end{eqnarray}
where $\eta_0$ has already been replaced by its value at horizon crossing.
This expression can then be used in \eqref{defcorrectedomega} to obtain the explicit form of the corrected
frequencies at first order in the slow-roll parameters. In this way, we can already
write our main equations, \eqref{eqomega0}
and \eqref{Omega1tildeq}, at this level of approximation.

\section{Uncorrected power spectra}

In this section, we will obtain the solution to Eq.~\eqref{eqomega0},
in order to construct the power spectra for
scalar and tensor perturbations at the level of approximation that
corresponds to the usual formalism of quantum 
field theory on classical background space-times. In
\cite{Martin:2012}, one can  
find a detailed computation in a formalism very similar to the present
one. 

Considering for now only scalar perturbations, the solution to Eq.~\eqref{eqomega0} with ${}^\text{S}\omega^2_{\vec{k}}$ given by \eqref{omegasgamma} can be written as follows:
\be \label{omegaeta}
{}^{\text{S}}\Omega_{\vec{k}}^{(0)}(\eta) = -\,\I\,\frac{{}^{\text{S}}y^{(0)\prime}_{\vec k}(\eta)}{{}^{\text{S}}y^{(0)}_{\vec k}(\eta)},
\ee
where the mode functions ${}^{\text{S}}y^{(0)}_{\vec k}$ can be
expressed in terms of the Bessel functions $J_\nu$: 
\begin{align}
{}^{\text{S}}y^{(0)}_{\vec k}=(-k\eta)^{1/2}
\bigl[&c_{{\vec k},1}\,J_{-(\gamma+3/2)}(-k\eta) \nonumber\\ \label{solbessel}
&+c_{{\vec k},2}\,J_{\gamma+3/2}(-k\eta)\bigr].
\end{align}
In order to obtain the Bunch-Davies vacuum for $\eta \rightarrow -\infty$, and choosing
the usual normalization of the Wronskian
\be
{}^{\text{S}}y^{(0)\prime}_{\vec k}\,{}^{\text{S}}y^{(0)*}_{\vec k} - {}^{\text{S}}y^{(0)\prime *}_{\vec k}\,{}^{\text{S}}y^{(0)}_{\vec k}=\I\,,
\ee
it is necessary to set
\begin{align}
\label{eq:bunchdaviesstandard}
c_{{\vec k},1}&=-c_{{\vec k},2}\,{\rm e}^{-\I\pi (\gamma +3/2)}, \\
c_{{\vec k},2}&=-\frac{\I}{2}\sqrt{\frac{\pi}{k}}\,\frac{{\rm e}^{-\I\pi/4+\I\pi(\gamma +3/2)/2}} 
{\sin[\pi (\gamma + 3/2)]}\,. \nonumber
\end{align}
In the super-Hubble limit $(-k\eta \rightarrow 0)$, the real part of ${}^{\text{S}}\Omega_{\vec{k}}^{(0)}(\eta)$
is given by
\be
\label{reomegaeta}
\Real {}^{\text{S}}\Omega^{(0)}_{\vec k}(\eta)=\frac{k\, \pi \, 2^{-2 (1 + \gamma)}}{\Gamma^2(\gamma+3/2)}
(-k\eta)^{2 (1 + \gamma)} \,.
\ee
Taking the inverse of this function and linearizing it in terms of the
slow-roll parameters, we get
the standard result for the power spectrum of the scalar perturbations
(see e.g. \cite{PU09}, p.~498): 
\begin{align}
{\cal P}_{\text{S}}^{(0)}(k) = 
\frac{G\,H^2_k}{\pi\,\epsilon}\bigl[1 -2\epsilon + 
\gamma(4 &- 2\gamma_\text{E} - 2\ln(2))\bigr],
\label{Ps0srk}
\end{align}
where $\gamma_\text{E} \simeq 0.5772$ is the Euler-Mascheroni
constant and the result should be evaluated 
at the horizon exit of the mode.

Up to a global multiplicative factor,
the power spectrum for the tensor modes can be immediately obtained
from the last expression by setting $\gamma = \epsilon$ for
the terms inside the square brackets, and reads 
\begin{align}
{\cal P}_{\text{T}}^{(0)}(k) = 
\frac{16\,G\,H^2_k}{\pi}\bigl[1 -2\epsilon + \epsilon(4 &-
2\gamma_\text{E} - 2\ln(2))\bigr]. 
\label{Pt0srk}
\end{align}
In the de~Sitter case, expression \eqref{omegaeta} simplifies
considerably to 
\be \label{om0ds}
{}^{\text{dS}}\Omega^{(0)}_{\vec{k}}(\eta):=\frac{k^3\eta^2}{1+k^2\eta^2}+\frac{{\rm i}}{\eta(1+k^2\eta^2)}.
\ee
(This corresponds to Eq. (129) in \cite{BKK}.)

Given the frequent appearance of the factor $-k\eta$, we now define the quantity
\be
\label{defxi}
\xi := -k\eta\,, 
\ee
which allows us to isolate the $k$-dependence of this expression:
\be \label{om0ds}
{}^{\text{dS}}\Omega_{\vec{k}}^{(0)}(\xi):=k\left[\frac{\xi^2}{1+\xi^2}-\frac{{\rm i}}{\xi(1+\xi^2)}\right]\,.
\ee
Note that the ${}^{\text{S}}\Omega_{\vec{k}}^{(0)}$, whose real part is given in
\eqref{reomegaeta}, only depends on the combination $\gamma$ of the
slow-roll parameters. 
For later convenience, we linearize it around its de~Sitter counterpart \eqref{om0ds},
\be\label{decomega0}
{}^{\text{S}}\Omega^{(0)}_{\vec{k}}={}^{\text{dS}}\Omega^{(0)}_{\vec{k}}+\gamma\,\Omega^{(0)}_{\gamma,\vec{k}}.
\ee
The term linear in the slow-roll parameters has the following form:
\be\label{om0gamma}
\Omega^{(0)}_{\gamma,\vec{k}}(\xi):=k\left[\frac{2 \xi^3 \E^{2 \I \xi} (\pi -\I \text{Ei}(-2 \I \xi ))-\xi^2-2 \I \xi-1}{\I\xi(\xi-\I)^2}\right];
\ee
\text{Ei} being the exponential integral function. In spite of the
appearance of this special function, this form of 
${}^{\text{S}}\Omega^{(0)}_\vec{k}$ is much more manageable than the
definition above in terms of derivatives of Bessel functions.

From this point on, we will skip the index $\vec{k}$ all along in order to
simplify the notation.

\section{Corrected power spectra}
\subsection{The corrected frequencies}

Inserting the form of the potential \eqref{Veps} into the definition (\ref{defcorrectedomega}) and linearizing in the slow-roll parameters,
we obtain the following form for the corrected frequencies for the scalar sector:
\be\label{omegadecomposition}
\widetilde\omega_\text{S}^2=\omega_\text{S}^2+\left(\frac{H_k^2}{m_\text{P}^2 \,k}\right)\Big(\widetilde\omega^2_\text{dS}+\epsilon\,\widetilde\omega^2_{\epsilon}
+\gamma\,\widetilde\omega^2_{\gamma} +\mathcal{O}(2)\Big);
\ee
the classical frequencies $\omega_\text{S}$ have been defined above in \eqref{omegasgamma}, while
the other terms stand for different quantum-gravity corrections and
are defined as follows: 
\begin{widetext}
\begin{eqnarray}
\widetilde\omega_{\text{dS}}^2&:=&\frac{\xi ^4 \left(\xi ^2-11\right)}{2 \left(\xi ^2+1\right)^3},\\\label{defomegaepsilon}
\widetilde\omega_{\epsilon}^2&:=&-\frac{\xi ^4}{6 \left(\xi ^2+1\right)^3}\left[
12 \left(\xi ^2+1\right)+\left(\xi ^2-11\right) \left(11-6 \ln\xi\right)\right],\\\label{defomegagamma}
\widetilde\omega_{\gamma}^2&:=&\frac{\xi ^4}{2 \left(\xi ^2+1\right)^4}\Big\{
2 \text{Ci}(2 \xi ) \left[2 \xi  \left(2 \xi ^6-9 \xi ^4+14 \xi ^2-11\right) \sin (2
   \xi )-\left(6 \xi ^8+10 \xi ^6+53 \xi ^4-12 \xi ^2+11\right) \cos (2 \xi )\right]\nonumber\\
   &&\qquad\qquad\qquad-7
   \xi ^6+21 \xi ^4-89 \xi ^2+2 \left(2 \xi ^6-9 \xi ^4+14 \xi ^2-11\right) \xi  \cos
   (2 \xi ) (\pi -2 \text{Si}(2 \xi ))\nonumber\\&&\qquad\qquad\qquad+\left(6 \xi ^8+10 \xi ^6+53 \xi ^4-12 \xi
   ^2+11\right) \sin (2 \xi ) (\pi -2 \text{Si}(2 \xi ))+27
\Big\}.
\end{eqnarray}
\end{widetext}
Since the difference between correction terms for the frequencies of the
tensorial and scalar modes are encoded in the $\Omega^{(0)}$ [see Eq.~\eqref{defcorrectedomega}],
the corrected frequencies for the tensor sector can be obtained directly
from (\ref{omegadecomposition}) by setting $\gamma=\epsilon$. Note
that, when written in 
terms of this dimensionless time variable $\xi$, the only explicit dependence 
 of the corrected frequencies on $k$ is encoded in the global factor
written in front of the quantum-gravity 
corrections in the decomposition \eqref{omegadecomposition}. Furthermore, $\text{Si}$ and $\text{Ci}$ are,
respectively, the sine and the cosine integral functions:
\begin{eqnarray}
\text{Si}(x):=\int_0^x\frac{\sin u}{u}\,\D u\,,\\
\text{Ci}(x):=-\int_x^\infty\frac{\cos u}{u}\,\D u\,.
\end{eqnarray}
Finally, the logarithmic term that appears in the definition of
$\widetilde\omega_{\epsilon}$ [see \eqref{defomegaepsilon}]
comes from the $\epsilon$ in the exponent in the form of the potential \eqref{Veps}.
Note that the argument of this logarithm should be $\xi/\xi_0$ but,
since $\xi_0$ is chosen as the Hubble-scale exit time, it turns out that $\xi_0=-k\eta_0=1$.

Let us comment now on the behavior of these corrected frequencies in the limit of late and early times.
At late times $\eta\rightarrow 0^-$, or equivalently
$\xi\rightarrow0^+$, it is possible to check 
by direct computation that
\be
\widetilde\omega_\text{S}^2=k^2-\frac{2+3\gamma}{\eta^2}+{\cal O}(\eta^4).
\ee
As can be seen, all the quantum-gravity corrections disappear in this limit.
On the other hand, for the limit at early times $\eta\rightarrow
-\infty$ ($\xi\rightarrow\infty$), 
one obtains that
\be\label{asymptoticsomega}
\widetilde\omega_\text{S}^2=k^2
+\frac{H_k^2}{2 k m_{\text{P}}^2}\left[1 +\frac{3}{2} \gamma 
-\frac{23}{3} \epsilon+2 \epsilon  \ln\xi
\right]+{\cal O}(1/\xi^2).
\ee
As commented in Sec.~III, this last result will provide us with the
natural initial data for $\Omega^{(1)}$ given in \eqref{initialdata} as well as for its linearized version $\widetilde\Omega^{(1)}$,
which is defined as $\widetilde\Omega^{(1)}=k-\widetilde\omega_\text{\text S}$ at $\xi\rightarrow\infty$.

\subsection{The linearized equation}

In order to solve the linear equation \eqref{Omega1tildeq} for
$\widetilde\Omega^{(1)}$, we proceed in a similar way as in the decomposition
\eqref{omegadecomposition} for the corrected frequency.
We split it into three different functions, each of them corresponding
to one kind of correction:
\be\label{decompositionomegatilde1}
\widetilde\Omega^{(1)}:=\left(\frac{H_k}{m_{\text{P}} \,k}\right)^2\left(\widetilde\Omega^{(1)}_{\text{dS}}
+\epsilon\,\widetilde\Omega^{(1)}_{\epsilon}
+\gamma\,\widetilde\Omega^{(1)}_{\gamma}\right),
\ee
where the overall prefactor has been written for convenience.
Recalling that $\widetilde\Omega^{(1)}$ is defined as
$\widetilde\Omega^{(1)}=\Omega^{(1)}-\Omega^{(0)}$
and that the natural initial data are given by $\Omega^{(1)}=\widetilde\omega$, at  $\xi\rightarrow\infty$,
the initial data corresponding to different objects defined above can be straightforwardly obtained
by linearizing the asymptotic behavior of the corrected frequency [the square root of the right-hand side
of \eqref{asymptoticsomega}]
on $H_k^2/m_\text{P}^2$:
\begin{eqnarray}\label{initialdata1}
\widetilde\Omega^{(1)}_{\text{dS}}&\xrightarrow[\xi\rightarrow\infty]{}&\frac{1}{4},\\\label{initialdata2}
\widetilde\Omega^{(1)}_{\epsilon}&\xrightarrow[\xi\rightarrow\infty]{}&-\frac{1}{12}\left[23-6\ln\xi\right],
\\\label{initialdata3}
\widetilde\Omega^{(1)}_{\gamma}&\xrightarrow[\xi\rightarrow\infty]{}&\frac{3 }{8}.
\end{eqnarray}
In the case of $\widetilde\Omega^{(1)}_{\epsilon}$, due to the presence of the logarithmic term,
the limit turns out to be divergent. This is obviously not a physical
consequence 
of the quantum-gravity corrections but an artifact of the slow-roll approximation
which, as mentioned in Sec. IV, is only valid for a large, but finite, interval of time.
Thus, in practice, the condition \eqref{initialdata2} will have to be imposed
for a  large, but finite, value of $\xi_{\text{initial}}\gg1$.

In order to find the equations of motion for the different $\widetilde\Omega^{(1)}$,
it is enough to insert the form \eqref{decompositionomegatilde1} into equation \eqref{Omega1tildeq},
linearize everything in the slow-roll parameters,
and collect terms with the same slow-roll coefficient (either 1, $\epsilon$
or $\gamma$). Following this procedure, Eq. \eqref{Omega1tildeq}
is rewritten as three different equations:
\begin{eqnarray}
-\,\I\,\frac{{\rm d}\widetilde\Omega^{(1)}_{\text{dS}}}{{\rm
    d}\xi}\,&=&\frac{2}{k}  \,\Omega^{(0)}_{\text{dS}}\,
\widetilde\Omega^{(1)}_{\text{dS}}-\widetilde\omega^2_{\text{dS}},
\label{eqdS}
\\\label{eqepsilon}
-\,\I\,\frac{{\rm d}\widetilde\Omega^{(1)}_{\epsilon}}{{\rm d}\xi}\,&=&\frac{2}{k} \,\Omega^{(0)}_{\text{dS}}\, \widetilde\Omega^{(1)}_{\epsilon}-\widetilde\omega^2_{\epsilon},\\\label{eqgamma}
-\,\I\,\frac{{\rm d}\widetilde\Omega^{(1)}_{\gamma}}{{\rm d}\xi}\,&=&\frac{2}{k} \,\Omega^{(0)}_{\text{dS}}\, \widetilde\Omega^{(1)}_{\gamma}
+ \frac{2}{k} \,\Omega^{(0)}_{\gamma}\, \widetilde\Omega^{(1)}_{\text{dS}}-\widetilde\omega^2_{\gamma},
\end{eqnarray}
where $\Omega^{(0)}_{\text{dS}}$ and $\Omega^{(0)}_{\gamma}$ have been
given above in Eqs.~\eqref{om0ds} and \eqref{om0gamma}. 
Equation \eqref{eqgamma} for $\widetilde\Omega^{(1)}_{\gamma}$ is coupled
to $\widetilde\Omega^{(1)}_{\text{dS}}$, whereas 
equation \eqref{eqepsilon} for $\widetilde\Omega^{(1)}_{\epsilon}$ is
independent of it. This is due to the presence
of the slow-roll parameter $\gamma$ in ${}^{\text{S}}\Omega^{(0)}$ \eqref{decomega0}.
Note that, due to the definitions and decompositions we have performed,
in particular \eqref{omegadecomposition} and \eqref{decompositionomegatilde1},
we have been able to write our fundamental equation
\eqref{Omega1tildeq} as three differential equations with initial data 
\eqref{initialdata1}--\eqref{initialdata3}, which do not
depend on any parameter: they only depend on the time variable $\xi$.
Neither $H_k$ nor $m_{\rm P}$ appear in these equations and also none of the slow-roll parameters. Furthermore,
we have written the equations in terms of the time variable $\xi$, such that there is no explicit dependence on $k$ either.
[Note that the explicit inverse of the $k$ factor that appears
multiplying $\Omega^{(0)}_{\text{dS}}$ and $\Omega^{(0)}_{\gamma}$ in these equations,
cancels exactly with their linear dependence on $k$, see Eqs.~\eqref{om0ds}--\eqref{om0gamma}].
Therefore, the explicit dependence of the result on the different parameters is analytically known.
In particular, one can already deduce
that the correction for the spectrum of the scalar perturbations
$\Delta_\text{S}$ defined in \eqref{defdeltas} will have
the following form:
\be \label{defdeltas2}
\Delta_\text{S}=\frac{H_k^2}{k^3 m_{\text{P}}^2}\Big[\beta_{\text{dS}}+\epsilon\,\beta_\epsilon + \gamma\,\beta_\gamma\Big],
\ee
with certain numerical factors $\beta_{\text{dS}}$, $\beta_\epsilon$
and $\beta_\gamma$. These factors
are given by using the decompositions \eqref{decomega0} and
\eqref{decompositionomegatilde1} 
in \eqref{defdeltas2} and linearizing in the slow-roll parameters,
\begin{eqnarray}\label{defbetads}
\beta_{\text{dS}}&=&-k \lim_{\xi\rightarrow 0}\left(\frac{\Real\widetilde\Omega^{(1)}_{\text{dS}}}{\Real\Omega^{(0)}_{\text{dS}}}\right),\\\label{defbetaepsilon}
\beta_{\epsilon}&=&-k \lim_{\xi\rightarrow 0}\left(\frac{\Real\widetilde\Omega^{(1)}_{\epsilon}}{\Real\Omega^{(0)}_{\text{dS}}}\right),\\
\beta_{\gamma}&=&k \lim_{\xi\rightarrow 0}\left(\frac{\Real\Omega^{(0)}_{\gamma}\Real\widetilde\Omega^{(1)}_{\text{dS}}
-\Real\Omega^{(0)}_{\text{dS}}\Real\widetilde\Omega^{(1)}_{\gamma}
}{\left(\Real\Omega^{(0)}_{\text{dS}}\right)^2}\right).\label{defbetagamma}
\end{eqnarray}
Note again that all these are $k$-independent  numbers. The explicit $k$ in front of the limit cancels out with the
global $k$ factors in the expressions of $\Real\Omega^{(0)}_{\text{dS}}$ and $\Real\Omega^{(0)}_{\gamma}$
[see Eqs.~\eqref{om0ds}--\eqref{om0gamma}].

The correction $\Delta_\text{T}$, corresponding to the tensorial
sector, can be obtained from the scalar one by imposing $\gamma=\epsilon$:
\be\label{defdeltat}
\Delta_\text{T}=\frac{H_k^2}{k^3 m_{\text{P}}^2}\Big[\beta_{\text{dS}}+\epsilon\,(\beta_\epsilon + \beta_\gamma)\Big].
\ee
Finally, the meaningful correction for the tensor-to-scalar ratio $r$ \eqref{tts0} will be given by the difference
between both:
\be
\label{deltatminusdeltas}
\Delta_\text{T}-\Delta_\text{S}=\frac{H_k^2}{k^3 m_{\text{P}}^2}(\delta-\epsilon)\,\beta_\gamma,
\ee
where we have reintroduced the second slow-roll parameter $\delta$.

In this subsection we have already achieved the principal goal of this paper; namely,
obtaining the specific forms of $\Delta_\text{S}$ \eqref{defdeltas2} and $\Delta_\text{T}$ \eqref{defdeltat}. The only issue
left is to obtain the values of the numerical coefficients \eqref{defbetads}--\eqref{defbetagamma}.
This will be performed in the rest of the present section by solving the three equations
\eqref{eqdS}--\eqref{eqgamma} above. For this purpose,  except for the de Sitter part,
which is analytically solvable, numerical simulations will have to be performed.

\subsection{The de Sitter part $\beta_{\text{dS}}$}

By construction, the equation for the de~Sitter part
$\widetilde\Omega^{(1)}_{\text{dS}}$ is the same 
as we found in our previous paper \cite{BKK} (see Eq. (144) there),
and it can be analytically solved. We will not repeat 
here the whole computation, but note that the behavior of the solution
at $\xi\rightarrow\infty$ is given by
\be
\widetilde\Omega^{(1)}_{\text{dS}}=\frac{1}{4}+C\,\E^{2 i \xi}+{\cal O}\!\left(\ln(\xi^{-2})\right).
\ee
In order to impose the initial condition as given by \eqref{initialdata1} and, in fact, to pick up
the only non-oscillating solution, the integration constant $C$ must be chosen
to be vanishing. After analyzing the behavior of the solution at the super-horizon limit,
one can show that the quantum-gravity correction term corresponding to the de Sitter
part takes the following specific value:
\be\label{valuebetads}
\beta_{\text{dS}}=\frac{1}{4 \E^2}\left[3 \E^4 \text{Ei}(-2)+9 \text{Ei}(2)-\E^2\right]\approx0.988,
\ee
cf.~Eq.~(147) in \cite{BKK}. 

\subsection{The $\epsilon$ part}

\begin{figure}
\includegraphics[width=0.5\textwidth]{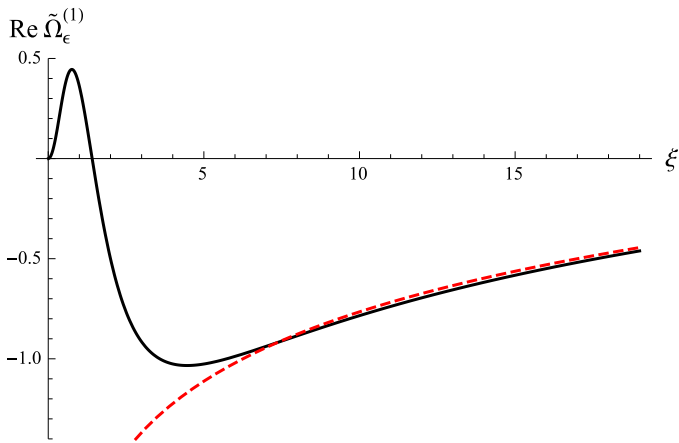}
\caption{The evolution of $\Real\widetilde\Omega^{(1)}_{\epsilon}$ (continuous black line)
and its asymptotic value (red dashed line).}\label{omega1epsilon}
\end{figure}

\begin{figure}
\includegraphics[width=0.5\textwidth]{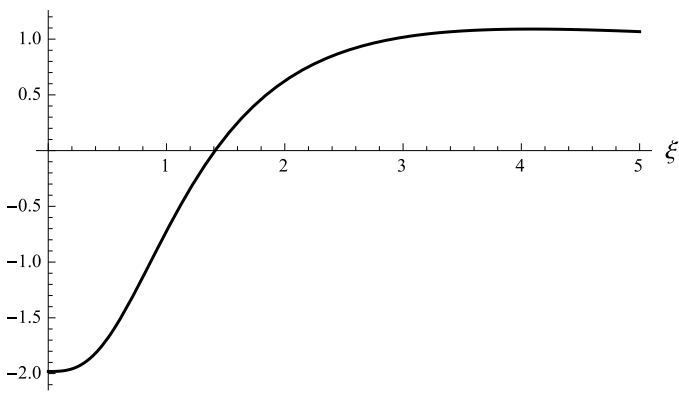}
\caption{The evolution of the expression that defines $\beta_\epsilon$ (the right-hand side
of Eq.~\eqref{defbetaepsilon} without taking the limit) as a 
function of $\xi$.}\label{betaepsilon}
\end{figure}

The difficulty in solving this part lies in the presence of the logarithmic term
both in the source \eqref{defomegaepsilon} and the asymptotic limit
\eqref{initialdata2}. In fact, 
if one removes this term, it is possible to solve equation \eqref{eqepsilon} and
obtain the value of $\beta_\epsilon$ analytically. This analysis
is presented in the Appendix, in order to show that the presence
of the logarithmic term, even if it is divergent in both early-time
($\xi\rightarrow\infty$) 
and late-time ($\xi\rightarrow 0$) limits, does not change the result
drastically. 

The complete equation \eqref{eqepsilon} must be solved by numerical
methods. In order to impose the initial condition \eqref{initialdata2},
one needs to choose an initial large value of $\xi=\xi_\text{initial}$
and insert that value into the function
\be
\widetilde\Omega^{(1)}_\epsilon(\xi_\text{initial})=
-\frac{1}{12}\left[23-6\ln\xi_\text{initial}\right].
\ee
Note that it is very convenient to pick up different values for
$\xi_\text{initial}$ to check that the final result does not depend on it. In particular,
we have chosen  $\xi_\text{initial}=10^6$, $10^7$, and $10^8$.
The absolute difference bet\-ween the various $\Omega^{(1)}_\epsilon(\xi)$,
as computed for those different values, is less than $10^{-6}$
during the whole evolution (for any of the considered values of $\xi$),
which translates to a negligible difference
of around $10^{-5}$ in the value of $\beta_\epsilon$.
In fact, in Fig.~\ref{omega1epsilon} it is possible to see
that $\Omega^{(1)}_\epsilon$ tends very quickly to its asymptotic
value, which explains the very weak dependence of the numerical
solution on the chosen value of $\xi_\text{initial}$.

Finally, in order to compute numerically $\beta_\epsilon$,
we have evaluated the expression that appears in its
definition \eqref{defbetaepsilon} for a very small value of $\xi=10^{-3}$,
which leads to
\be\label{valuebetaepsilon}
\beta_\epsilon\approx-1.98.
\ee
Note that all the different $\Omega$ tend to vanish as
$\xi\rightarrow 0$ and thus one cannot numerically compute
the limit \eqref{defbetaepsilon} by evaluating it at $\xi=0$. Nonetheless,
in Fig.~\ref{betaepsilon} the evolution of the right-hand side of \eqref{defbetaepsilon}
(without the limit taken)
has been plotted, in order to show that the limit behaves smoothly
and is well defined.

\subsection{The $\gamma$ part}

Equation \eqref{eqgamma} for $\widetilde\Omega^{(1)}_{\gamma}$ is the
most intricate one. 
It is, in particular, coupled to the equation for
$\widetilde\Omega^{(1)}_{\text{dS}}$. As in the previous subsection, for the
computation of $\beta_\epsilon$,
there seems to be no way to write the solution analytically in terms of special
functions and thus, it is necessary to resort to a numerical resolution.
Nevertheless, the presence of the sine and cosine integral functions in the source term \eqref{defomegagamma}
makes it computationally highly demanding to deal with this equation numerically
for large values of their argument. Thus, at early times ($\xi\gg 1$), it will be very convenient to consider
instead their series expansion.

Therefore, we have applied the following procedure. First, these two functions
are replaced by their series expansion in inverse powers of $\xi$ up to 20th order
in the source term \eqref{defomegagamma}.
Then, equation \eqref{eqgamma} is solved with this approximated source term and by imposing
the initial data $\widetilde\Omega^{(1)}_{\gamma}=3/8$ for a very large initial time $\xi=\xi_\text{initial}\gg1$.
Next, the obtained numerical solution is used as initial data
at a still large, but smaller value, of
$\xi=\xi_\text{intermediate}<\xi_\text{initial}$, for Eq.~\eqref{eqgamma}
with the full form of its source term \eqref{defomegagamma}.
Finally, this latter equation is solved and, with this numerical solution at hand,
one can compute the expression on the left-hand side of Eq. \eqref{defbetagamma}, whose limit
as $\xi\rightarrow 0$ will define our quantity of interest $\beta_\gamma$.
As in the previous subsection, this limit is numerically computed by evaluating
the corresponding expression for a small value of $\xi$.

In order to check the robustness of this method, different values for $\xi_\text{initial}$ and $\xi_\text{intermediate}$
have been used. In particular, we have chosen $\xi_\text{initial}$ as
$10^8$, $10^9$, and $10^{10}$, while
for $\xi_\text{intermediate}$ the values 100, 400, and 800 have been picked up.
We find that the largest absolute difference between $\Real\widetilde\Omega^{(1)}_{\gamma}$,
as computed with the solution corresponding to these different values, during the whole evolution
turns out to be smaller than $10^{-4}$.
This is translated to an absolute difference of a similar order for $\beta_\gamma$.
In fact, a value of $\beta_\gamma\approx 2.56$ is found and, thus, the error due to assuming
an approximate equation for early times is very small.

\begin{figure}
\includegraphics[width=0.5\textwidth]{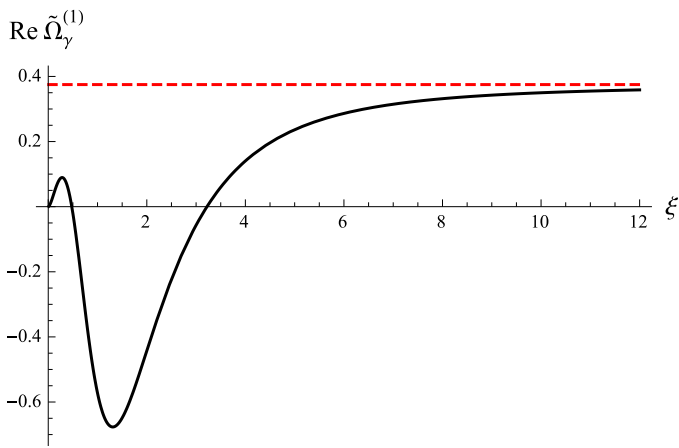}
\caption{The evolution of $\Real\widetilde\Omega^{(1)}_{\gamma}$ (black continuous line)
and its asymptotic value 3/8 (red dashed line).}\label{omega1gamma}
\end{figure}

\begin{figure}
\includegraphics[width=0.5\textwidth]{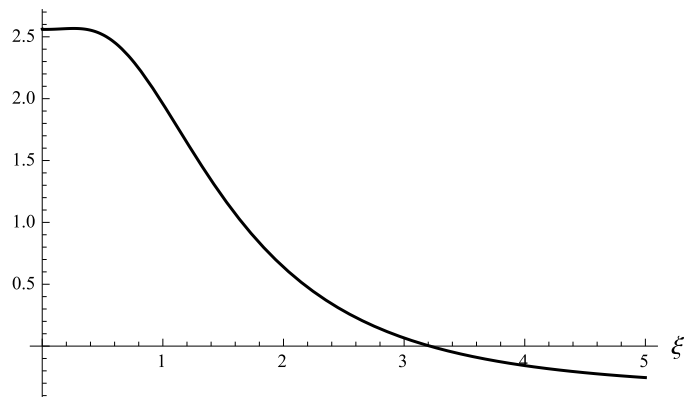}
\caption{The evolution of the expression that defines $\beta_\gamma$ (the right-hand side
of Eq.~\eqref{defbetagamma} without taking the limit) as a function of $\xi$.}\label{betagamma}
\end{figure}

The evolution of the real part of $\widetilde\Omega^{(1)}_{\gamma}$ is shown
in Fig. \ref{omega1gamma}, in combination with its asymptotic value $3/8$.
Note that $\widetilde\Omega^{(1)}_{\gamma}(\xi)$ is a non-oscillating
solution and approaches its asymptote from below very quickly.
In Fig.~\ref{betagamma},
the evolution of the relation that defines $\beta_\gamma$ [the right-hand side
of relation \eqref{defbetagamma} without taking the limit] is shown in terms of $\xi$.
As can be seen clearly in the figure, its late-time limit behaves
smoothly and leads to the value
\be\label{valuebetagamma}
\beta_\gamma\approx2.56.
\ee

\section{Results and observability}

In this section we will summarize our results and obtain the explicit form of the
different parameters of the power spectrum, in particular the spectral indices and their running.
We will also comment on the magnitude of the obtained corrections and the possibility
of observing them. Finally, we will give the form of the correction for the $C_\ell$ coefficients used in the CMB data analysis.

\subsection{Parameters of the power spectra}

The main results of this paper are the corrected forms of the power spectra for
gauge-invariant scalar and tensor modes: 
\begin{eqnarray}
{\cal P}_{\text{S}}^{(1)}(k) &=&
 {\cal P}_{\text{S}}^{(0)}(k)\Big\{1+\Delta_{\text{S}}\Big\}\,,\\
{\cal P}_{\text{T}}^{(1)}(k) &=& {\cal P}_{\text{T}}^{(0)}(k)\Big\{1+\Delta_{\text{T}}\Big\}\,,
\end{eqnarray}
where ${\cal P}_{\text{S}}^{(0)}(k)$ and ${\cal P}_{\text{T}}^{(0)}(k)$ are the usual
power spectra, which can be obtained in the approximation of quantum
fields propagating 
on fixed cosmological backgrounds and which are explicitly given in
\eqref{Ps0srk} and \eqref{Pt0srk}, 
respectively. These correspond to the standard result derived usually by
other means (see e.g.~\cite{PU09}).
The quantum-gravity corrections are thus encoded in the
$\Delta_\text{S}$ \eqref{defdeltas2} 
and $\Delta_\text{T}$ \eqref{defdeltat} terms, with the
$\beta_{\text{dS}}$,  $\beta_{\epsilon}$ 
and  $\beta_{\gamma}$ computed, respectively, in \eqref{valuebetads},
\eqref{valuebetaepsilon}, and \eqref{valuebetagamma}.
Inserting explicitly these numerical values and replacing the auxiliary parameter $\gamma$ defined in \eqref{gammadef} by $\epsilon$ and $\delta$, the corrections to the power spectra read as follows:
\begin{eqnarray}
\Delta_\text{S}&=&
 \frac{H_k^2}{m_{\text{P}}^2}\left(\frac{\bar{k}}{k}\right)^3\bigl(0.988+3.14\,\epsilon - 2.56\,\delta\bigr)\,, \label{Ps1}\\
\Delta_\text{T}&=& \frac{H_k^2}{m_{\text{P}}^2}\left(\frac{\bar{k}}{k}\right)^3\bigl(0.988+0.58\,\epsilon\bigr)\,. \label{Pt1}
\end{eqnarray}
Here, we have reverted the rescaling applied in \eqref{repl1} and a reference wave number $\bar{k}=1/\mathfrak{L}$
has been defined as the inverse of the length scale introduced to regularize the spatial integral.

Since the dependence of the quantum-gravity corrections on the $k$ wave numbers
have been analytically obtained, it is possible to derive the spectral
indices and their runnings by direct computation.
We will use the usual parametrization of the power spectra given by the following power-law relation,
\begin{eqnarray}\label{parametrization1}
{\cal P}_{{\text{S}}}^{(1)}(k) &=&A_{{\text{S}}} \left(\frac{k}{k_*}\right)^{n_\text{S}-1+\alpha_\text{S} \ln(k/k_*)},\\\label{parametrization2}
{\cal P}^{(1)}_{{\text{T}}}(k) &=&A_{{\text{T}}} \left(\frac{k}{k_*}\right)^{n_\text{T}+\alpha_\text{T} \ln(k/k_*)},
\end{eqnarray}
where higher-order terms in $\ln(k/k_*)$ have been neglected in the exponent and the pivot scale $k_*$ has been introduced. 
In this way, at a linear level in the slow-roll parameters,
we obtain for the scalar sector
\begin{align}
&n_\text{S}-1 :=\left.\frac{{\rm d} \ln {\cal P}_{\text{S}}^{(1)}(k)}{{\rm d}\ln k}\right|_{k=k_*} \\
&= 2\delta -4 \epsilon-\frac{H_k^2}{m_\text{P}^2}\left(\frac{\bar{k}}{k_*}\right)^3\bigl[
3\beta_{\text{dS}}+\epsilon(2\beta_{\text{dS}}+3\beta_\epsilon+6\beta_\gamma) \nonumber\\
& \qquad\qquad\qquad\qquad\qquad\qquad\qquad\qquad\qquad\qquad-3\delta\beta_\gamma 
\bigr]\nonumber \\
&\approx 2\delta -4 \epsilon-\frac{H_k^2}{m_\text{P}^2}\left(\frac{\bar{k}}{k_*}\right)^3\left(
2.96+11.40\,\epsilon-7.68\,\delta\right), \nonumber
\end{align}
with the first two terms taken together being the usual first-order approximation of the spectral index.
In order to obtain this result, the relation
\be
\frac{{\rm d} \ln H_k}{{\rm d} \ln k}=-\epsilon,
\ee
has been used. 
Similarly, for the tensor sector, one obtains
\begin{eqnarray}
n_\text{T} &:=&\left.\frac{{\rm d} \ln {\cal P}_{\text{T}}^{(1)}(k)}{{\rm d}\ln k}\right|_{k=k_*} \\
&=& -\,2\epsilon-\frac{H_k^2}{m_\text{P}^2}\left(\frac{\bar{k}}{k_*}\right)^3\left[
3\beta_{\text{dS}}+\epsilon(3\beta_\epsilon+3\beta_\gamma+2\beta_{\text{dS}})
\right] \nonumber \\
&\approx& -\,2\epsilon-\frac{H_k^2}{m_\text{P}^2}\left(\frac{\bar{k}}{k_*}\right)^3\left(2.96+3.72\,\epsilon\right). \nonumber
\end{eqnarray}
Finally, one can also obtain the running of the spectral indices,
\begin{eqnarray}
\alpha_\text{S}:=\left.\frac{{\rm d}n_\text{S}}{{\rm d}\ln k}\right|_{k=k_*},\\
\alpha_\text{T}:=\left.\frac{{\rm d}n_\text{T}}{{\rm d}\ln k}\right|_{k=k_*},
\end{eqnarray}
by taking into account the dependence of the slow-roll parameters
on the wave number. In particular, up to second-order terms we have
\begin{eqnarray}
\frac{{\rm d}\epsilon}{{\rm d}\ln k}&=&2 \epsilon (\epsilon-\delta),\\
\frac{{\rm d}\delta}{{\rm d}\ln k}&=&2 \epsilon (\epsilon-\delta)-\theta,
\end{eqnarray}
where the second-order slow-roll parameter $\theta$ is defined as
\be
\theta:=\frac{\dot\epsilon-\dot\delta}{H}.
\ee
In this way, one gets by direct computation the results
\begin{eqnarray}
\alpha_\text{S}&=&4\epsilon(\delta-\epsilon)-2\theta\\
& & \!\!+\,\frac{3 H_k^2}{m_\text{P}^2}\left(\frac{\bar{k}}{k_*}\right)^3\!
\bigl[3\beta_{\text{dS}}+\epsilon
  (4\beta_{\text{dS}}+3\beta_\epsilon+6\beta_\gamma)- 3\delta\beta_\gamma
\bigr] \nonumber\\
&\approx&4\epsilon(\delta-\epsilon)-2\theta \nonumber\\
& &\!\!+\,\frac{
  H_k^2}{m_\text{P}^2} \left(\frac{\bar{k}}{k_*}\right)^3
\left(8.89+40.12\,\epsilon - 23.04\,\delta\right), \nonumber\\
\alpha_\text{T}&=&4\epsilon(\delta-\epsilon)\\ & &\!\!+\,\frac{3 H_k^2}{m_\text{P}^2}\left(\frac{\bar{k}}{k_*}\right)^3
\Bigl[3\beta_{\text{dS}}+\epsilon
  (4\beta_{\text{dS}}+3\bigl(\beta_\epsilon+\beta_\gamma)\bigr) \Bigr] \nonumber \\
  &\approx&4\epsilon(\delta-\epsilon)+\,\frac{
  H_k^2}{m_\text{P}^2} \left(\frac{\bar{k}}{k_*}\right)^3
\left(8.89+17.08\,\epsilon\right), \nonumber
\end{eqnarray}
where only the first-order slow-roll terms have been kept in the
correction term. Note that the prefactors in the correction term for
the runnings get larger than for the spectral indices due to the fact
that the usual parametrization \eqref{parametrization1}--\eqref{parametrization2}
is not a good fit to the obtained quantum-gravity correction.

Let us finally also give the quantum-gravitational correction to the
$r$-parameter \eqref{tts0}. Since this correction is given by
\eqref{deltatminusdeltas}, we obtain
\be\label{scalartensorratio}
r^{(1)}\approx 16\epsilon\left(1+2.56\,\frac{H_k^2}{m_{\rm
      P}^2} \left(\frac{\bar{k}}{k}\right)^3(\delta-\epsilon)\right). 
\ee
In most models $\delta-\epsilon=-\dot{\epsilon}/2H\epsilon<0$,
thus the correction gives a small negative contribution to
the scalar-to-tensor ratio. Note that, in the pure de~Sitter case
there is no correction for this quantity, so it arises entirely
from the slow-roll part.

\subsection{Estimation of the magnitude of the correction}

In order to give an estimation of the magnitude of the corrections for different
quantities, one necessarily needs to assume a specific value for the scale $\bar{k}=1/\frak{L}$.
It seems reasonable to understand this scale as an infrared cut-off and
relate it to the largest scale that could influence the CMB ($\bar{k}\approx 10^{-4}\,\text{Mpc}^{-1}$).
Nonetheless, in the literature there are some analyses that try to fix its
value by other means. In particular, in Ref.~\cite{KTV16}, which is based on
a similar semiclassical approach to the geometrodynamical quantization
of the present problem, arbitrary parameters are considered in
front of the correction factor $H_k^2 \bar{k}^3/(m_{\rm P}^2k^3)$ and
they are fitted with the observational data from the Planck mission \cite{Planck}.
In particular, the best fit obtained relates $\bar{k}$ to the size of
galaxies or galaxy clusters ($\bar{k}\approx 1\,\text{Mpc}^{-1}$). The meaning of such
a scale is, however, not clear. 
All the above being said, in order to give an estimate of the magnitude
of the correction we have obtained, here we will assume
$\bar{k}$ to be equal to the pivot scale chosen by the Planck mission,
that is $\bar{k}=k_*=0.05\,\text{Mpc}^{-1}$.
At the end of this subsection, we will comment on the maximum
possible value of $\bar{k}$ not to contradict the experimental data.

As we have shown in our previous work \cite{BKK}, taking into account the energy scale of inflation
in combination with the upper bound given by the Planck mission \cite{Planck} for the scalar-to-tensor ratio
$r \lesssim 0.11$, it is possible to derive the following condition,
\be\label{maxvalueHmp}
\frac{H_{\text{inf}}}{m_{\rm P}} \lesssim 1.3 \times 10^{-5}\,,
\ee
$H_{\text{inf}}$ being the average Hubble parameter during inflation.
In the following, in order to give estimates for the correction to the quantities derived in the previous section, we will assume that $H_k=H_{\text{inf}}$.

In addition, the experimental constraint for the spectral index, $n_\text{S} \approx 0.968\pm0.006$ (see \cite{Planck}),
implies that $\epsilon \lesssim 0.007$ and $\delta \approx -0.002$. With these values at hand, we can give an estimate
for the upper limits of the quantum-gravity correction for scalar and tensor perturbations as follows:
\begin{eqnarray}
\left|\Delta_{\text{S}}\right|  &\lesssim& 2 \times 10^{-10},\,\quad\left|\Delta_{\text{T}}\right|  \lesssim 2 \times 10^{-10}.
\end{eqnarray}
Since the value of the slow-roll parameters is so small and the dominant de Sitter contribution $\beta_\text{dS}$ has a value close to 1, the approximated upper bounds for the corrections
coincide with the approximated maximal value of the ratio $H_k^2/m_{\rm P}^2$.
Using the estimated values for the slow-roll parameters, we can, however, deduce that the corrections for both kinds of perturbations differ by about $2\,\%$:
\be
\frac{\Delta_{\text{S}}}{\Delta_{\text{T}}} \approx 1.02\,.
\ee
Inserting the estimated numbers for $\epsilon$ and $\delta$, we can immediately see that the correction to the spectral index is significantly smaller than the statistical uncertainty in the Planck data:
\be
\left[n_\text{S}^{(1)} - n_\text{S}^{(0)}\right]_{k=\bar{k}} \approx - 3.1\,\frac{H_{\text{inf}}^2}{m_\text{P}^2} \approx -5 \times 10^{-10}\,.
\ee
In this case, the quantum-gravitational correction to the spectral index is also tiny,
\be
\left[n_\text{T}^{(1)} - n_\text{T}^{(0)}\right]_{k=\bar{k}} \approx - 3.0\,\frac{H_{0,\text{inf}}^2}{m_\text{P}^2} \approx -5 \times 10^{-10}\,.
\ee
Estimating the magnitude of the quantum-gravity correction for the running gives
\begin{align}
\left[\alpha_\text{S}^{(1)} - \alpha_\text{S}^{(0)}\right]_{k=\bar{k}} &\approx 9.2\,\frac{H_{\text{inf}}^2}{m_\text{P}^2} \approx 2 \times 10^{-9}\,, \\
\left[\alpha_\text{T}^{(1)} - \alpha_\text{T}^{(0)}\right]_{k=\bar{k}} &\approx 9.0\,\frac{H_{\text{inf}}^2}{m_\text{P}^2} \approx 2 \times 10^{-9}\,.
\end{align}
Finally, the upper bound for the correction of the scalar-to-tensor ratio can be estimated as
\be
\left[\frac{r^{(1)}-r^{(0)}}{r^{(0)}}\right]_{k=\bar{k}} \approx -0.023\,\frac{H_{\text{inf}}^2}{m_\text{P}^2} \approx -4 \times 10^{-12}\,.
\ee
As can be seen, all corrections are very small and they are inside the current experimental error bars.

Let us finally comment on the maximally allowed value for $\bar{k}$ by the experimental data.
The fact that the experimental errors are larger than the corrections \eqref{Ps1}--\eqref{Pt1} we have obtained
leads to the following relation,
\be
{\bar{k}}_{\rm max}=\left(\frac{m^2_{\rm P} \Delta_{\rm exp}}{H^2_{\rm inf}}\right)^{1/3} k_*,
\ee
$\Delta_{\rm exp}$ being the relative experimental error in the power spectrum.
In order to give a rough estimate, we assume that this error is of order one, which, as can be seen
in \cite{HS15}, is a very high bound. Using the maximum value for the ratio $H_{\rm inf}/m_{\rm P}$
found in \eqref{maxvalueHmp}, we get
\be \label{kmax}
{\bar{k}}_{\rm max}\approx 100\, {\rm Mpc}^{-1}.
\ee
This implies a minimum value for the length scale ${\mathfrak L}_{\rm min}\approx 10^{-2}\, {\rm Mpc}$.
Note, even though, that a lower value of $H_{\rm inf}/m_{\rm P}$ would increase ${\bar{k}}_{\rm max}$.

\subsection{CMB temperature anisotropies}

In this subsection we will obtain the correction for the CMB
temperature anisotropies, which are usually expressed by the quantities $C_\ell$ as defined below. In order to obtain these coefficients, it is
necessary to evolve the scalar power spectrum through subsequent phases of
the universe from the end of inflation until today. In addition, one finally
needs to project it on the celestial sphere. This whole procedure can be
reduced to computing the following integral, which is usually done
numerically,
\be
C^{(i)}_\ell=\int_{0}^{\infty} \frac{\D k}{k}\,\mathcal{P}^{(i)}_\text{S}(k)\,\Theta_{\ell}^2(k),
\ee
with $i=0,1$ denoting the uncorrected and corrected coefficients, respectively, and $\Theta_{\ell}(k)$ being the transfer function.
For large scales (small $\ell$), however, it is possible to
solve this integral analytically. In this regime, the fluctuations were well outside the
horizon at the end of recombination and thus they were not affected by subhorizon
physics. Therefore, it is only necessary to take into account the primordial spectrum
and perform the projection on the celestial sphere.
In particular, the transfer function can in this case be given in terms of the spherical Bessel
functions $j_\ell$ as follows \cite{Dodelson},
\be
\Theta_{\ell}(k)=\frac{1}{3}\,j_\ell(k[\eta_\text{hor}-\eta_{\text{rec}}]),
\ee
where $\eta_\text{hor}$ is the conformal time at horizon crossing and $\eta_{\text{rec}}$
the conformal time at recombination.

Let us define the quantum-gravitational correction to the temperature anisotropies in the following
way,
\be
\Delta C_\ell:=C_\ell^{(1)}- C_\ell^{(0)}.
\ee
Applying the results for the corrected scalar power spectrum, we get that,
for large scales, this correction has the following form,
\be
\Delta C_\ell\approx \frac{1}{4\pi^2}\int_{0}^\infty \frac{{\rm d}k}{k\,\epsilon}\left(\frac{H_k}{m_\text{P}}\right)^4\left(\frac{\bar{k}}{k}\right)^3\, j^2_\ell(k[\eta_\text{hor}-\eta_{\text{rec}}]),
\ee
The approximate symbol in this equation stands for two reasons. On the one
hand, we are assuming an approximated transfer function. And, on the other
hand, the overall factor that appears in the correction term \eqref{Ps1},
which depends on the slow-roll parameters but is of order one, has been dropped.
At this point, as explained in \cite{Baumann}, we use the fact that the Bessel function
is strongly peaked around $k|\eta_\text{hor}-\eta_\text{rec}|\approx \ell$ and effectively acts as a Dirac
delta mapping between $k$ and $\ell$. Therefore, one can integrate the explicit $k$
dependences in the last integral, which leads to
\be
\Delta C_\ell\approx\!\!\frac{3}{4\pi\epsilon}\!\!\left(\!\frac{H_k}{m_\text{P}}\!\right)^4\!\!\!\!
\frac{|\bar{k}(\eta_\text{hor}-\eta_{\text{rec}})|^3}{(2\ell-3)(2\ell-1)(2\ell+1)(2\ell+3)(2\ell+5)},
\ee
while the implicit $k$-dependences on $H_k$ and $\epsilon$ should
be evaluated at $k|\eta_\text{hor}-\eta_{\text{rec}}|\approx \ell$.
Applying exactly the same approximations, it is straightforward to obtain
also the well-known result for the uncorrected temperature anisotropies
at large scales:
\be
C_{\ell}^{(0)}\approx\frac{1}{8\pi^2\epsilon} \left(\!\frac{H_k}{m_\text{P}}\!\right)^2\frac{1}{\ell(\ell+1)}\,,
\ee
which does not have any explicit dependence on $(\eta_0-\eta_{\rm rec})$ and,
for a scale-invariant spectrum (constant $H_k$ and $\epsilon$) is just proportional
to the inverse of $\ell(\ell+1)$.

\begin{figure}
\includegraphics[width=0.5\textwidth]{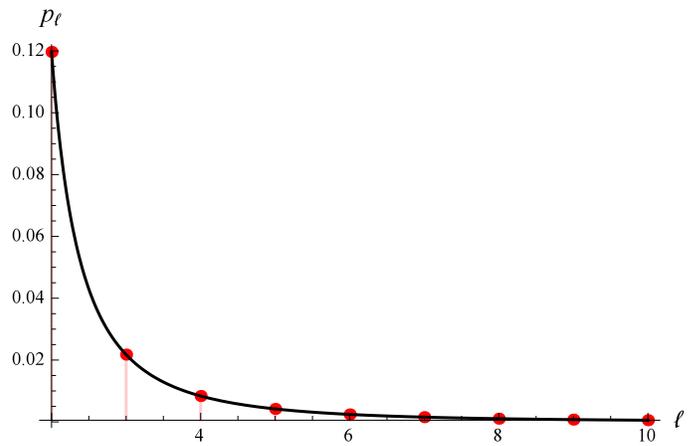}
\caption{The ratio of the correction $\Delta C_\ell$ to the uncorrected $C_{\ell}^{(0)}$ without the non-numerical prefactors, such that we have \\ $p_\ell := 3\pi\,\ell(\ell+1)\left[(2\ell-3)(2\ell-1)(2\ell+1)(2\ell+3)(2\ell+5)\right]^{-1}$.}\label{pl-plot}
\end{figure}

Thus, due to a correction proportional to
$k^{-3}$ in the power spectrum, the temperature anisotropies get
a correction that goes as the inverse of a fifth-order polynomial in $\ell$
for large scales. However, since the uncorrected $C_\ell^{(0)}$ go with $\ell^{-2}$, the relative correction $\Delta C_\ell/C^{(0)}_\ell$ is of order $\ell^{-3}$.
As commented in the previous section, especially due to the presence of
$\bar{k}$ is this correction, it is difficult to be certain about its absolute value.
We have thus plotted the behavior of the relative correction with respect to $\ell$ without the physical prefactors in Fig.~\ref{pl-plot}.
We can also check the relative values and see that the correction
drops very quickly with $\ell$. For instance, comparing it for the first
multipoles:
\be\nonumber
\frac{\Delta C_{\ell=3}}{\Delta C_{\ell=2}}\approx 0.09,\quad\frac{\Delta C_{\ell=4}}{\Delta C_{\ell=2}}\approx 0.02,
\quad\frac{\Delta C_{\ell=5}}{\Delta C_{\ell=2}}\approx 0.007.
\ee
It is important to note that the qualitative behavior derived in this section
for correction of the temperature anisotropies essentially comes from
the explicit $k^{-3}$ dependence of the correction of the power spectrum,
which has been obtained in several approaches (see, for instance, \cite{KTV16}).

We can also give an estimate of how large the Hubble parameter, i.e.~the energy scale, during inflation would have to be such that one could see a correction of our type. Given that cosmic variance behaves like
\be
\frac{\Delta C_\ell^{^\text{CV}}}{C_\ell^{(0)}} =\sqrt{\frac{2}{2\ell +1}},
\ee
one can conclude that for $\ell = 2$, where cosmic variance is about $\Delta C_2^{^\text{CV}}/C_2^{(0)} \approx 0.63$ and our quantum-gravitational correction is given by
\be
\frac{\Delta C_2}{C_2^{(0)}} \approx 0.12\,\left(\!\frac{H_k}{m_\text{P}}\!\right)^2\,|\bar{k}(\eta_\text{hor}-\eta_{\text{rec}})|^3,
\ee
the remaining factors in the above expression would have to be larger than 5 in order to clearly
see an effect in the CMB data. Using $\bar{k} =0.05\,\text{Mpc}^{-1}$, and given that
$|\eta_\text{hor}-\eta_{\text{rec}}|$ can be estimated to be about $700\,\text{Mpc}$
(see e.g.~Tab.~I in \cite{AG16}), the factor $|\bar{k}(\eta_\text{hor}-\eta_{\text{rec}})|^3$
turns out to be of order $5\times10^{4}$. Therefore, in order to see an effect,
$(H_k/m_\text{P}) \gtrsim 10^{-2}$ would be required, which is by far 
outside the range \eqref{maxvalueHmp}
allowed by the measured tensor-to-scalar ratio. Moreover, if we use the
maximum allowed value by the latter limit for the Hubble factor, it would
be necessary for $\bar{k}$ to be around $\bar{k} \approx 5\,\text{Mpc}^{-1}$
to get an observable effect. This value is significantly smaller than the
maximum value ${\bar{k}}_{\rm max}$ derived in \eqref{kmax}.

\section{Conclusions}

In this paper quantum-gravitational corrections for the power spectra of
the gauge-invariant scalar and tensor
perturbations have been obtained in the slow-roll regime.
In particular, the corrected form for the scalar and tensor power spectrum
is given, respectively, by \eqref{PS0omega} and \eqref{PT0omega},
where $\Delta_\text{S}$ is given in \eqref{defdeltas2} 
and $\Delta_\text{T}$ in \eqref{defdeltat} in terms of the numerical coefficients
$\beta_{\text{dS}}$,  $\beta_{\epsilon}$ and  $\beta_{\gamma}$. 
These coefficients have been computed by solving the
linearized evolution equation for the Gaussian width $\Omega^{(1)}$ with
natural initial data (in the sense that the initial state, which is constructed as
a small deformation of the usual Bunch-Davies vacuum, best describes in this context the
expected properties of a freely evolving mode). Their values are given
in \eqref{valuebetads}, \eqref{valuebetaepsilon}, and \eqref{valuebetagamma},
respectively.

The above results generalize the results for the de~Sitter case
obtained earlier in \cite{BKK} to a more realistic scenario of slow-roll
inflation. In particular, and as one would naively expect, the main part
of the correction is due to the de Sitter contribution (which introduces
an enhancement of the spectrum), whereas the slow-roll
part slightly modifies it. Let us at this point
briefly comment on the results of \cite{KTV16}, obtained from an
alternative expansion of the Wheeler-DeWitt equation. As in our
treatment, the authors find
quantum-gravitational correction terms proportional to
$H_\text{inf}^2/(m_{\rm P}^2k^3)$. Nonetheless,
their result for the slow-roll approximation is not just a small perturbation
of the de~Sitter case, but can give a comparable contribution for
large scales, which can even lead to a power loss instead of an
enhancement of the power spectra.

Moreover, let us stress that
the kind of correction that has been obtained in this analysis -- being
proportional to the factor $H_\text{inf}^2 \bar{k}^3/(m_{\rm P}^2k^3)$ --
has appeared in several different approaches in the context of quantum
geometrodynamics \cite{Sasha1,Sasha2,Sasha3,KTV16}.
The form of this correction is not completely unexpected.
In fact, it is possible to argue, already on dimensional grounds, that $(H_\text{inf}/m_{\rm P})^2$
is the only non-dimensional parameter that one could use
to include perturbatively (as a power series expansion) quantum-gravity
corrections. Furthermore since, due to the background homogeneity,
one needs to introduce explicitly a volume (${\mathfrak L}^3=1/\bar{k}^3$)
in order to regularize the spatial integral in the action, another dimensionless
quantity $(\bar{k}/k)^3$ enters the game.
Nevertheless, in principle, the power of this latter quantity might have been different
and thus it is very interesting to see how the same correction is explicitly realized
in different specific models.

In the last section we have also analyzed the magnitude of the obtained corrections
and the possibility of observing them experimentally. The most difficult issue
in order to give a precise estimation is that, due to the regularization of
the spatial integral in the action, a length scale needs to be considered.
The power spectrum then depends on that length scale and there seems
to be nothing physical to fix it. As we have commented, in the main part
of the paper, the most reasonable choice is to take it as an infrared cut-off,
relating it to the largest observable scale in the CMB. In our case,
just to give an approximated estimate, we have chosen it as
the length scale of a typical mode that affects the CMB. In particular
we have chosen the pivot scale selected by the Planck mission.
In this way, we have obtained that the corrections for all different parameters
of the power spectra (spectral indices and runnings) are well inside
the current experimental error bars.

Finally, we have also obtained the qualitative form of the correction
induced in the CMB temperature anisotropies by this quantum-gravity effect.
The analysis we have performed
is valid for large scales (small $\ell$), for which quantum-gravity
effects are expected to be more relevant. In particular, it shows that
a correction of the form $k^{-3}$ which, as commented above,
seems very generic in this context, leads to a relative correction of
the order $\ell^{-3}$ for the anisotropies, which thus quickly declines
with increasing $\ell$.

With this paper we conclude our investigations on quantum-gravitational corrections arising from a canonical quantization of a perturbed universe model using the Wheeler-DeWitt equation. The effects on large scales we have obtained are, for a reasonable choice of $\bar{k}$, not observable in the CMB data and since we have used a generic slow-roll model that encompasses a wide range of inflationary models, using more refined models that obey the slow-roll approximation would not enhance the corrections.
Nonetheless, it is still an open question whether such corrections can be observed in
situations where cosmic variance is not present; for example, in galaxy-galaxy
correlation functions.

\section*{Acknowledgments}

We thank I\~naki Garay and Jon Urrestilla for interesting discussions.
D.\,B.~is supported by Project No.~FIS2014-57956-P of the Spanish Ministry
of Economy and Competitiveness and by Project No.~IT956-16 of the Basque Government. The research of M.\,K.~was financed
by the Polish National Science Center Grant DEC-2012/06/A/ST2/00395. 
This article is based upon work from COST Action CA15117 ``Cosmology and Astrophysics Network for Theoretical Advances and Training Actions (CANTATA)", supported by COST (European Cooperation in Science and Technology).

\appendix

\section{Computation of $\beta_\epsilon$ removing the divergent logarithmic term}

\begin{figure}
\includegraphics[width=0.5\textwidth]{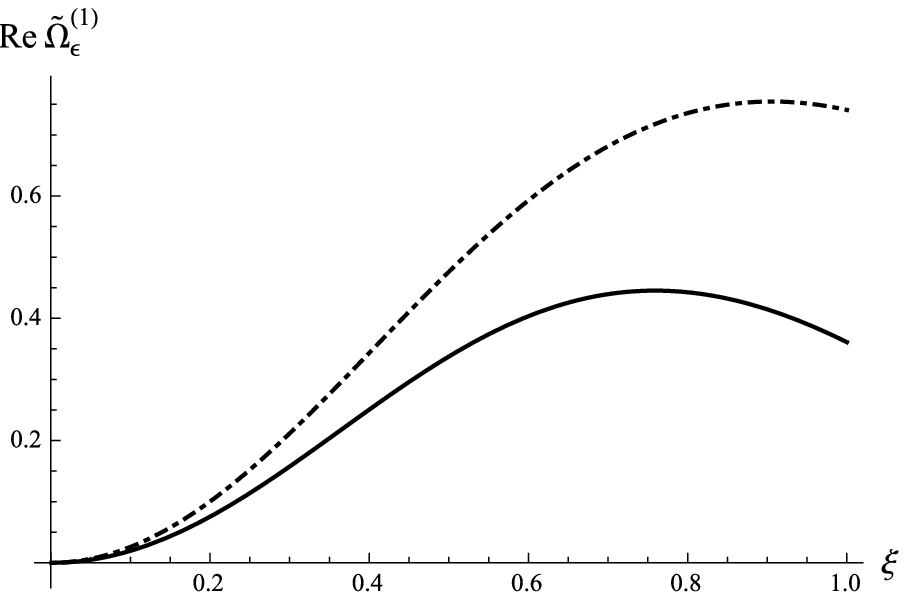}
\includegraphics[width=0.5\textwidth]{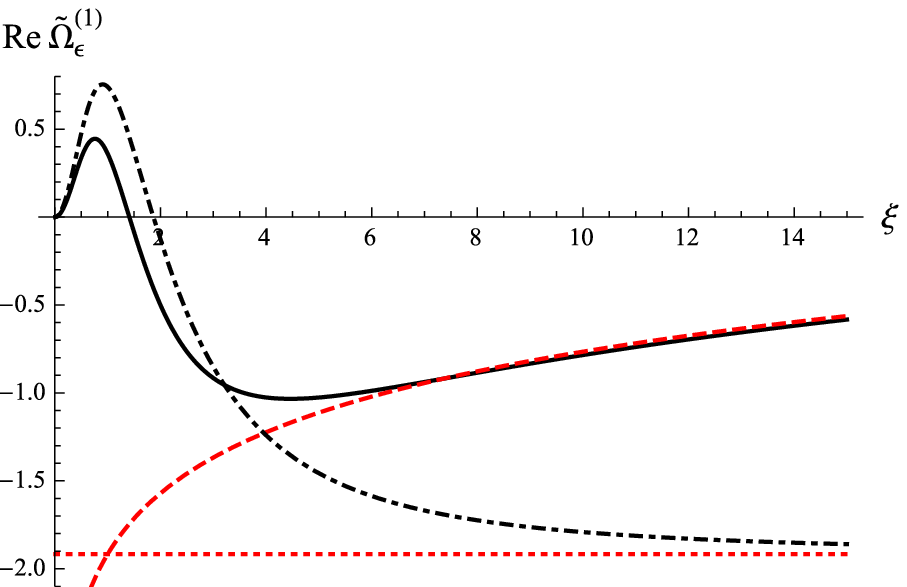}
\caption{The evolution of the real part of $\widetilde\Omega^{(1)}_\epsilon$
is shown, for different ranges of values of $\xi$, as given by its full equation
(black continuous line) and as obtained
by dropping the logarithmic term (black dashed-dotted line). Their corresponding
asymptotes are shown as a red-dashed line and as a red-dotted line,
respectively.} \label{comparisonomega1epsilon}
\end{figure}

In this Appendix, Eq.~\eqref{eqepsilon}, with the logarithmic term dropped,
will be solved in order to show that, even if this term is divergent in both
late and early time limits, the value that is obtained for $\beta_\epsilon$
does not change dramatically. 
Note that, if one removes the logarithmic term from the source term \eqref{defomegaepsilon},
one should also remove it from the initial condition \eqref{initialdata2}.

Interestingly, in this case Eq.~\eqref{eqepsilon} can be analytically solved,
and the solution takes the following form:
\begin{align}
\widetilde\Omega^{(1)}_\epsilon =\;&\frac{1}{12}\,\E^2 \xi ^2 \left(\xi ^2+1\right)^2
\Big[3\,\E^{2 {\rm i} \xi } (\xi +\I)^2 (-4 \E^2 C \nonumber\\
&+33 \, \Gamma (0,2 \I \xi -2) +11 \,\E^4 \Gamma (0,2 \I \xi +2)) \nonumber\\
&-\E^2 \left(12 \left(\xi^2+1\right)+11 \,(1+\xi  (\xi -6\I))\right)\Big],
\end{align}
where $C$ is an integration constant. If we analyze the behavior of this
solution at $\xi\rightarrow\infty$, we find that
\be
\widetilde\Omega^{(1)}_\epsilon\approx-\frac{23}{12}-C\,\E^{2 \I \xi}.
\ee
Therefore, in order to have a non-oscillating solution, we choose $C=0$.

Finally, we compute the limit defined in Eq.~\eqref{defbetaepsilon}
and find the following value of $\beta_\epsilon$:
\be
\beta_\epsilon=1+\frac{11}{12\,\E^2}(\E^2-3 \,\E^4 \text{Ei}(-2)-9\,\text{Ei}(2))\approx-2.62.
\ee
As commented above, this proves that the logarithmic term is indeed important
to compute the precise value for $\beta_\epsilon$, but it is not
critical in the sense 
that qualitatively the same result is obtained if one drops it.

In Fig.~\ref{comparisonomega1epsilon}, the evolution of the real part
of $\widetilde\Omega^{(1)}_\epsilon$ is shown for both the solution
with and without the logarithmic term, in combination with their
corresponding asymptotes. It can be seen that the tendency
at late times is quite similar for both, which explains the
weak dependence of $\beta_{\epsilon}$ on the commented term.

\end{document}